\documentclass[12pt]{article}

\usepackage{authblk}
\usepackage{amssymb}
\usepackage{amsmath}
\usepackage{mathrsfs}
\usepackage{amsfonts}
\usepackage{braket}
\usepackage{color}
\usepackage{graphicx}
\usepackage{tikz}
\usetikzlibrary{shapes,snakes}
\usepackage{pgfplots}
\pgfplotsset{compat=1.9} 
\newtheorem{thm}{Theorem}

\newtheorem{lem}[thm]{Lemma}
\newtheorem{prop}[thm]{Proposition}

\newtheorem{defn}[thm]{Definition}
\newcommand{\norm}[1]{\left\Vert#1\right\Vert}

\def\XXint#1#2#3{{\setbox0=\hbox{$#1{#2#3}{\int}$ }
\vcenter{\hbox{$#2#3$ }}\kern-.6125\wd0}}

\newcounter{lastnote}

\title{Covariant Ergodic Quantum Markov Semigroups via Systems of Imprimitivity}
\author{Radhakrishnan Balu}
\affil{Army Research Laboratory Adelphi, MD, 21005-5069, USA \\
       radhakrishnan.balu.civ@mail.mil
      }  
\affil{Department of Mathematics \&
	Norbert Wiener Center for Harmonic Analysis and Applications, 
	University of Maryland,
	College Park, MD 20742.\\
	rbalu@umd.edu}          
\date{Received: date / Accepted: \today}
\date{Received: date / Accepted: \today}
\begin{document}
\maketitle

\begin{abstract}
 We construct relativistic quantum Markov semigroups from covariant completely positive maps. We proceed by generalizing a step in Stinespring{'s} dilation to a general system of imprimitivity and basing it on Poincar\`e group. The resulting noise channels are relativistically consistent and the method is applicable to any fundamental particle, though we demonstrate it for the case of light-like particles. The Krauss decomposition of the relativistically consistent completely positive identity preserving maps (our set up is in Heisenberg picture) enables us to construct the covariant quantum Markov semigroups that are uniformly continuous. We induce representations from the little groups to ensure the quantum Markov semigroups that are ergodic due to transitive systems imprimitivity.
\end{abstract}

\section{Introduction}
\label{intro}
 Quantum channels that are described by completely positive unital maps are important ingradients in quantum information processing (QIP). Quantum Markov semigroups (QMS) that are composed of completely positive maps describe open quantum systems that form the basis for investigating quantum optics circuits \cite {Heinz-Peter2007}, \cite {Tezak2012}, \cite {Rolando2006} and quantum filtering \cite {Belavkin2005} among other examples where the interactions with a bath are taken into account. The interacting bath can be explicitly described by an infinite set of quantum harmonic oscillators leading to quantum stochastic differential equations \cite {KP1992} or using quantum Markov semigroups by tracing out the environmental degrees of freedom leading to Lindbladian quantum master equations. In both situations the Markovianity of the dynamics is assumed and the noise channels in terms of Krauss operators are obtained. In the QMS description discrete noise channels with quantum Poisson statistics will be absent. The noise channels in relativistic quantum quantum information (RQI) \cite {Terno2006} can be investigated within the framework of covariant QMS which is the focus of this work.

QMS that are covariant with respect to a group action has been studied by Fagnola et al. \cite {Fagnola2020} with several results on invariant subspaces obtained specific to compact groups. In quantum information processing invariant subspaces are important as they provide pathways for coherent evolution of quantum systems to ensure fidelity of quantum computation or communication. In this work our group of interest is the locally compact Poincar\`e relevant in relativistic quantum physics and the QMS is constructed from covariant completely positive unital maps. In our earlier work we treated covariant free quantum fields via Lorentz Group Representation of Weyl Operators\cite {Rad2020} constructed fock spaces for massless Weyl fermions and covariant white noises. The relativistically consistent quantum white noises \cite {Hida1977} are a rigorous description of noises of an open quantum system. In this work we focus on the system degrees of freedom along with the channels at the level of QMS that can indeed be dilated to the description of quantum stochastic differential equations or trajectories as is known in the physics community.
 
Systems of imprimitivity \cite{Mackey1963}, \cite {Varadarajan1985}, \cite {WignerLz1939} is a compact characterization of dynamical systems, when the symmetry of the kinematics of a quantum system is described by a group, from which infinitesimal forms in terms of differential equations ($Shr\ddot{o}dinger$, Heisenberg, and Dirac etc), localizability, and the canonical commutation relations can be derived \cite {Wigner1949} and \cite {Rad2019}. For example, the kinematics of a finite dimensional quantum system can be characterized up to an unitary isomorphism by a transitive system of imprimitivity with the underlying symmetry described by the cyclic group $Z_n$ where n is the dimension of the system \cite {Tolar2014}.  In the context of quantum information processing the Pauli Z operator can be considered as the position observable and the conjugate momentum-like opertaor is the Pauli X operator. Our important tool is Mackey machinery that characterizes such systems in terms of induced representations, specifically from little groups, applied to Poincar\`e group. 

To build covariant dilations of completely positive maps we break up steps of the proof of Stinespring theorem into construction of a map from the kernel induced by the completely positive map, a derived *-unital homomorphism and an isometry composite, and an unitary isomorphism to an extended Hilbert space. The last step was accomplished by considering a system of imprimitivity, based on the Poincar\'e group, endowed with an homogeneous projection valued measure (PVM) and by deriving a PVM of higher multiplicity. The Krauss operators are constructed from the unitary operators, that are irreducible representations of the group, and the isometry. By chaining the completely positive maps as a semigroup we obtain a covariant QMS. When the SI is transitive we get an ergodic QMS. Most of the steps are standard and our contribution in selecting the transitive SI, by inducing representations from little groups, to guarantee an ergodic QMS.

\section {Little groups (stabilizer subgroups)}
Some of the different systems of imprimitivity that live on the orbits of the stabilizer subgroups are described below. It is good to keep in mind the picture that SI is an irreducible unitary representation of Poincar$\grave{e}$ group $\mathscr{P}^+$ induced from the representation of a subgroup such as $SO_3$, which is a subgroup of homogeneous Lorentz, as $(U_m(g)\psi)(k) = e^{i\{k,g\}}\psi(R^{-1}_mk)$ where g belongs to the $\mathscr{R}^4$ portion of the Poincar$\grave{e}$ group, m is a member of the rotation group, and the duality between the configuration space $\mathbb{R}^4$ and the momentum space $\mathbb{P}^4$ is expressed using the character the irreducible representation of the group $\mathbb{R}^4$ as:
\begin {align} \label {eq: poissonBr}
\{k,g\} &= k_0 g_0 - k_1 g_1 - k_2 g_2 - k_3 g_3, p \in \mathbb{P}^4. \\
\hat{p}:x &\rightarrow e^{i\{k,g\}}. \\
\{Lx, Lp \} &= \{ x, p \}. \\\
\hat{p}(L^{-1}x) &= \hat{Lp}(x).
\end {align}
In the above L is a matrix representation of Lorentz group acting on $\mathbb{R}^4$ as well as $\mathbb{P}^4$ and it is easy to see that $p \rightarrow Lp$ is the adjoint of L action on $\mathbb{P}^4$. The $\mathbb{R}^4$ space is called the configuration space and the dual $\mathbb{P}^4$ is the momentum space of a relativistic quantum particle.

The stabilizer subgroup of the Poincar$\grave{e}$ group $\mathscr{P}^+$ (Space-like particle) can be described as follows: \cite {Kim1991}:
The Lorentz frame in which the walker is at rest has momentum proportional to (0,1,0,0) and the little group is
again SO(3, 1) and this time the rotations will change the helicity. 
\
\section {Fiber bundle representation of relativistic quantum particles}

The states of a freely evolving relativistic quantum particles are described by unitary irreducible representations of Poincar$\grave{e}$, which is a semidirect product of two groups, group that has a geometric interpretation in terms of fiber bundles. 

\begin {defn}
Let A and H be two groups and for each $h\in{H}$ let $t_{h}:a\rightarrow{h[a]}$ be an automorphism (defined below) of the group A. Further, we assume that $h\rightarrow{t_h}$ is a homomorphism of H into the group of automorphisms of A so that
\begin {align} \label{semidirectEq}
h[a] &= hah^{-1}, \forall{a\in{A}}. \\
h &= e_H, \text{  the identity element of H}. \\
t_{{h_1}{h_2}} &= t_{h_1}t_{h_2}.
\end {align}
Now, $G=H\rtimes{A}$ is a group with the multiplication rule of $(h_1,a_1)(h_2,a_2) = (h_1{h_2},a_1{t_{h_1}}[a_2])$. The identity element is $(e_H,e_A)$ and the inverse is given by $(h,a)^{-1} = (h^{-1},h^{-1}[a^{-1}]$. When H is the homogeneous Lorentz group and A is $\mathbb{R}^4$ we get the Poincar$\grave{e}$ group $\mathscr{P}=H\rtimes{A}$ via this construction. The covering group of inhomogeneous Lorentz is also a semidirect product as $\mathscr{P}^* = H^*\rtimes \mathbb{R}^4$ and as every irreducible projective representation of $\mathscr{P}$ is uniquely induced from a representation of $\mathscr{P}^*$ we will work with the covering group, whose orbits in momentum space are smooth, in the following.
\end {defn}
We will need the following lemma for our discussions on constructing induced representations using characters of an abelian group as in the case of equation \eqref {eq: poissonBr}.
\begin {lem} (Lemma 6.12 \cite {Varadarajan1985})
Let $h \in H$. Then, $\forall x \in \hat{A}$ where $\hat{A}$ is the set of characters of the group A (which in our case is $\mathbb{R}^4$), there exists one and only $y \in \hat{A}$ such that $y(a) = x(h^{-1}[a]), \forall a \in A$. If we write y = h[x], then $h,x \rightarrow h[x]$ is continuous from $H \times \hat{A}$ into $\hat{A}$ and $\hat{A}$ becomes a H-space. Here, y can be thought as the adjoint for action of H on $\hat{A}$ and the map $\hat{p}$ in equation\eqref {eq: poissonBr} is such an example that is of interest to our constructions. In essence, we have Fourier analysis when restricted to the abelian group A of the semidirect product.
\end {lem}

\section {Smooth Orbits in Momentum Space}
We need to described few ingredients to construct the Hilbert space of space-like fermions namely, the fiber bundle, the fiber vector space, an inner product for the fibers, and an invariant measure. 
The 3+1 spacetime Lorentz group $\hat{O}(3,1)$-orbits  of the momentum space $\mathscr{R}^4$, where the systems of imprimitivity established will live, described by the symmetry $\hat{O}(3,1)\rtimes\mathbb{R}^4$. The orbits have an invariant measure $\alpha^+_m$ whose existence is guaranteed as the groups and the stabilizer groups concerned are unimodular and in fact it is the Lorentz invariant measure $\frac{dp}{p_0}$ for the case of forward mass hyperboloid.  The orbits are defined as:
\begin {align}
Y^{+,1/2}_{im} &= \{p: p^2_0 - p^2_1 - p^2_2 - p^2_3 = -m^2, p_0 > 0\}, \text{\color{blue} single sheet hyperboloid}.
\end {align}
Each of these orbits are invariant with respect to $\hat{O}(3,1)$ and let us consider the stabilizer subgroup of the first orbit at p=(0,1,0,0). Now, assuming that the spin of the particle is 1/2 and mass $m > 0$  let us define the corresponding fiber bundles (vector) for the positive mass hyperboloid that corresponds to the positive-energy states by building the total space as a product of the orbits and the group $SL(2, C)$.
\begin {align} \label {eq: bundle}
\hat{B}^{+,1/2}_{im} &= \{(p,v) \text{   }p\in{\hat{Y}^{+1/2}_{im}, }\text{   }v\in\mathscr{C}^4,\sum_{r = 0}^3 p_r \gamma_r v = mv\}. \\
\hat{\pi} &: (p,v) \rightarrow {p}. \text{  Projection from the total space }\hat{B}^{+,2}_{im} \text{ to the base }\hat{Y}^{+,1/2}_{im}.
\end {align}
It is easy to see that if $(p, v) \in B_{im}^{+,1/2}$ then so is also $(\delta(h)p, S(h^{*-1})v)$. Thus, we have the following Poincar$\grave{e}$ group symmetric action on the bundle that encodes spinors into the fibers:
\begin {equation}
h,(p,v) \rightarrow (p,v)^h = (\delta(h)p, S(h^{*-1})v).
\end {equation}

We define the states of the light like particles \cite {Rad2020} on the Hilbert space $\hat{\mathscr{H}}^{+,1/2}_0$, square integrable functions on Borel sections of the bundle $\hat{B}^{+,1}_0 = \{(p,v) : (p,v) \in B_0^+, \Gamma v = \mp v\}$.

The states of the particles are defined on the Hilbert space $\hat{\mathscr{H}}^{+,1/2}_m$, square integrable functions on Borel sections of the bundle $\hat{B}^{+,2}_m$ with respect to the invariant measure $\beta^{+,1}_0$, whose norm induced by the inner product is given below:
\begin {equation} \label {eq: section}
 \norm{\phi}^2 = \int_{X^+_m}p_0^{-1}\langle\phi{p},\phi{p}\rangle {d\beta}^{+,1}_0(p).
\end {equation}
The invariant measure and the induced representation of the Poincar\'e group from that of the Weyl fermion are given below:
\begin {align}
{d\beta}^{+,1}_0(p) &= \frac {dp_1 dp_2 dp_3} {2(p_1^2 + p_2^2 + p_3^2)}. \\
(U_{h,x}\phi)(p) &= \text{exp i}\{x, p\} \phi (\delta(h)^{-1}p)^h.
\end {align}

\section {Stinespring dilation}
\begin {thm} \cite {Stinespring1955} Let $\mathscr{H}_1, \mathscr{H}_2$ be Hilbert spaces and $T: \mathscr{B}(\mathscr{H}_2)\rightarrow \mathscr{B}(\mathscr{H}_1)$ be a linear operator satisfying the following conditions: \\
(1) T(1) = 1, $T(X^*) = T(X)^*$; (ii) $\norm {T(X)} < \norm{X}$; (iii) if $X_n \rightarrow X$ weakly in $\mathscr{H}_2$ then $T(X_n) \rightarrow T(X)$ weakly in $\mathscr{H}_1$; (iv)$\forall n = 1,2 , \dots$ the correspondence $((X_{ij})) \rightarrow ((T(X_{ij}))), 1 \leq i, j \leq n$ from $\mathscr{B}(\mathscr{H}_1 \otimes \mathbb{C}^n))$ into $\mathscr{B}(\mathscr{H}_2 \otimes \mathbb{C}^n))$ preserves positivity.

Then, there exists a Hilbert space k, an isometry V from $\mathscr{H}_1$ into $\mathscr{H}_2 \otimes k$ such that (a) $T(X) = V^* X \otimes 1V, \forall X \in \mathscr{B}(\mathscr{H}_2)$; (b) $\{ (X \otimes 1)Vu | X \in \mathscr{B}(\mathscr{H}_2), u \in \mathscr{H}_1 \}$ is total in $\mathscr{H}_2 \otimes k$; (c) This association is valid up to an unitary isomorphism.
\end {thm}
The following definitions and the proof, we will elaborate the steps of one proposition that we want to focus up on, of Stinespring theorem are from the text book by Parthasarathy \cite {KP1992}.

\textbf{Positive definite kernel:} Let $\mathscr{H}$ be any set, possibly countably infinite. A positive definite kernel on $\mathscr{H}$ is a complex number valued map $K:\mathscr{H}\times$$\mathscr{H}\rightarrow$$\mathscr{C}$ satisfying
\begin{equation}
\sum\limits_{i,j}\bar{\alpha_i}\alpha_j{K(x_i,x_j)}\geq{0}, \alpha_i\in\mathscr{C},x_i\in\mathscr{H}.
\end{equation}

\begin {prop}  Let $T: \mathscr{B}(\mathscr{H}_2)\rightarrow \mathscr{B}(\mathscr{H}_1)$ be a completely positive map. Let us define a kernel
\begin {align}
K(X_1, Y_1, u_1; X_2, Y_2, u_2) &= \langle u_1, Y_1^*T(X_1^*X_2)Y_2u_2\rangle, \\ &\forall (X_i, Y_i, u_i) \in \mathscr{B}(\mathscr{H}_2) \times \mathscr{B}(\mathscr{H}_1) \times \mathscr{H}_1, i=1,2.
\end {align}
Then, K is a positive definite kernel on $\mathscr{B}(\mathscr{H}_2) \times \mathscr{B}(\mathscr{H}_1) \times \mathscr{H}_1$
\end {prop}
Next, we extract a map $\lambda$ that is part of the Gelfand pair via the following proposition.
\begin {prop} 
Let $T: \mathscr{B}(\mathscr{H}_2)\rightarrow \mathscr{B}(\mathscr{H}_2)$ be a completely positive map. Then, there exists a Hilbert space $\mathscr{H}$ and a map $\lambda$ such that the following holds up to an unitary: \\
(i) $\{\lambda(X, u)|X \in \mathscr{B}(\mathscr{H}_2, u \in \mathscr{H}_1\}$ is total in $\mathscr{H}$; \\
(ii) $\lambda$ is linear in each of the variables X, u; \\
(iii) $\langle \lambda(X_1, u_1), \lambda(X_2, u_2) \rangle = \langle u_1, T(X_1^*, X_2)u_2 \rangle$; \\
(iv) In particular the map $V_0 : u \rightarrow \lambda(1, u)$ is an isometry from $\mathscr{H_1}$ into $\mathscr{H}$.
\end {prop}
This leads to the existence of a homomorphism to the bath Hilbert space that is vital in synthesizing the noise operators.
\begin {prop}
Let $T, \lambda, V_0$ be as above. Then, there exists a *-unital homomorphism $\pi : \mathscr{B}(\mathscr{H}_2) \rightarrow \mathscr{B}(\mathscr{H})$ such that $\lambda(X, u) = \pi(X)V_0 u,\forall X \in \mathscr{B}(\mathscr{H}_2), u \in \mathscr{H}_1$.
\end {prop}
The next step, that decomposes the bath space operators via Hahn-Hellinger theorem, is our focus and so let us state and look at the details of the proof.
\begin {prop} {Dilation step:}
Let $\pi$ a *-unital homomorphism $\pi : \mathscr{B}(\mathscr{H}_2) \rightarrow \mathscr{B}(\mathscr{H})$. There exists,a Hilbert space k and an unitary isomorphism $\Gamma : \mathscr{H} \rightarrow \mathscr{H}_2 \otimes k$ such that $\pi(X) = \Gamma^{-1} X \otimes 1\Gamma, \forall X \in \mathscr{B}(\mathscr{H}_2)$ where 1 is the identity operator in k. The algebra generated by  U, V is $\mathscr{B}(\mathscr{H}_2)$ and linear combinations of $U_s V_t, s,t \in \mathbb{R}$ are weakly dense in $\mathscr{B}(\mathscr{H}_2)$ it follows that $\pi(X) = \Gamma^{-1}X \otimes 1\Gamma$.
\end {prop}

Let us consider only the case when dim $\mathscr{H}_2 = n < \infty$ and refer the readers to the above reference for the infinite case. Then up to unitary equivalence there exists a unique irreducible pair U, V of unitary operators in $\mathscr{H}_2$ such that $U^n = V^n = 1$ and $UV = VU exp \frac {2\pi i}{n}$, that is they obey Weyl commutator relation.
By analogue of the Stone-von Neumann theorem for the group $Z_n$ there exists an Hilbert space k such that $\Gamma^{-1}U \otimes 1\Gamma = \pi(U), \Gamma^{-1}V\otimes 1\Gamma = \pi(V).$ We get the noise channels as $Vu = L_1 u \oplus L_2 u \oplus \dots \oplus L_m u, m \leq n -1$. It is easy to verify that the dimension of k is bounded by $n_1 n_2$ that form the ancilla qubits in quantum information processing.

Let us reformulate the Weyl commutator relation in terms of SI \cite {Tolar2014} with
$\{U_g | g \in Z_n \}$ as the unitary representation of the group in the Hilbert space $\mathscr{H}$, and
\begin {align*}
\mathscr{E} &= \{ E(S), S \text{ is Borel subset of M }\}. \\
M &= \{ 0, 1, \dots, n - 1 \}. \\
U_g E(S) U_g^{-1} &= E(g^{-1}S).
\end {align*}
It is easy to see that the above unitary representation is induced from that of the subgroup $H = \{ 0 \}$ of $Z_n$.

To get the usual picture of a system interacting with a bath let us consider the space $h_0 \otimes \mathscr{H}$. We can define an unitary isomorphism $W: h_0 \otimes \mathscr{H} \rightarrow h_0 \oplus h_0 \oplus \dots$ as $Wu \otimes v = \oplus \langle f_i, v \rangle u$ where $\{ f_i \}$ forms the orthonormal basis of $h_0$. Then, the operator L can be viewed as acting on the system Hilbert space as $Lu = \bigoplus_j L_j u, u \in h_0$.

We need a way to unitarily transport an SI relation from an arbitrary Hilbert space to one that is a disjoint union of countably many (infinite or finite) of them that in turn is mediated by transporting a homogeneous projection valued measure \cite {Varadarajan1985}. We then apply it to Poincar\`e group to build the covariant completely positive maps and QMS.

\begin {defn} Let $\mathscr{H} = \mathscr{L}^2 (X, \mathscr{K}, \alpha)$ and $P = P(\mathscr{K}, \alpha)$ be a projection valued measure (PVM) $E \rightarrow P_E$ where $P_E f = \chi_E f, f \in \mathscr{H}$. The PVM acting on a separable Hilbert space is said to be homogeneous if it is equivalent to a $P(\mathscr{K}, \alpha)$ for suitable $\mathscr{K}$ and $\alpha$. The dimension of $\mathscr{K}$ is the multiplicity of P. Given any measure class $\mathscr{C}$ on X and any integer $1 \leq n \leq \infty$ there is a unique PVM P with $\mathscr{C}$ as the measure class of P and n its multiplicity.
\end {defn}
\begin {thm} Let (U, P) be a system of imprimitivity acting in $\mathscr{H}$ such that P is homogeneous of multiplicity $1 \leq n \leq \infty$. Let $\alpha$ be a $\sigma$-finite measure in the measure class $\mathscr{C}_P$ of P. Then (U, P) is equivalent to a system $(U{'}, P^{n, \alpha})$ acting in $\mathscr{H}_{n, \alpha}$.
\end {thm}
Let us state and prove the main result on covariant QMS.
\begin {prop} Let $T: \mathscr{B}(\hat{\mathscr{H}}^{+,1/2}_0) \rightarrow \mathscr{B}(\hat{\mathscr{H}}^{+,1/2}_0)$ be a completely positive map. Then, T can be dilated into a covariant map $V: \mathscr{B}(\hat{\mathscr{H}}^{+,1/2}_0) \rightarrow \mathscr{B}(\hat{\mathscr{H}}^{+,1/2}_0)\otimes \mathscr{B}(k)$. A relativistically consistent QMS can be constructed by composing $V_t$ parameterized by time.
\end {prop}
\textbf{(Sketch of proof):} In the dilation step of the Stinespring theorem we need to transport an SI relation that can be guaranteed if the PVM involved is a homogeneous one. This is the case for finite dimensional Hilbert spaces, as there is only one SI up to an unitary transformation, and for the infinite case we can ensure it by starting with a transitive system of imprimitivity. This can be achieved by inducing a representation from a closed subgroup which is light-like little group in our case as follows:

The states of the particles are defined on the Hilbert space \cite {Rad2020} $\hat{\mathscr{H}}^{+,2}_0$, square integrable functions on Borel sections of the bundle $\hat{B}^{+,2}_0$ with respect to the invariant measure $\beta^{+,1}_0$, whose norm induced by the inner product is given below:
\begin {equation} \label {eq: section}
 \norm{\phi}^2 = \int_{X^+_m}p_0^{-1}\langle\phi{p},\phi{p}\rangle {d\beta}^{+,1}_0(p).
\end {equation}
The invariant measure and the induced representation of the Poincar\'e group from that of the Weyl fermion are given below:
\begin {align}
{d\beta}^{+,1}_0(p) &= \frac {dp_1 dp_2 dp_3} {2(p_1^2 + p_2^2 + p_3^2)}. \\
(U_{h,x}\phi)(p) &= \text{exp i}\{x, p\} \phi (\delta(h)^{-1}p)^h.
\end {align}
The representation $(U_{h,x}\phi)$ that gives rise to a transitive SI can be transported across the *-unital homomorphism $\pi$ in the dilation step. The transitivity of the system ensures that it is also ergodic. It is important to note that in the case of infinite dimensional systems more than one dilation is possible for a given completely positive map and we have realized the dilation based on light-like particles. The construction of a covrainat QMS can proceed in the usual manner as a one parameter semigroup of the above completely positive map T. Suppose T has the Krauss operator representation as
\begin {equation}
T(X) = \sum_j L^*_j X L_j, \sum_j L^*_j L_j = I. 
\end {equation}
then the generator of the QMS is given by 
\begin {equation}
\Theta(X) = -\frac {1}{2} \lambda \sum_j L^*_j L_j X + XL^*_j L_j -2 L^*_j X L_j. 
\end {equation}

\section {Summary and conclusions}
We derived the covariant QMS using Mackey{'s} machinery on systems of imprimitivity. We dilated a completely positive operator by first building a transitive system of imprimitivity and transported across an unitary at the dilation step of the Stinespring construction. Our construction based on light-like particle is one of several possible ways to dilate a given completely positive map. We continued with the construction of a covriant QMS with the usual approach of gluing together completely positive maps as a semigroup. Our work on QMS covariant with respect to the actions by locally compact groups that are relevant in RQI opens up opportunities for extending invariant subspace results of compact groups. 


\end{document}